\documentclass[a4paper]{article}
\usepackage{amsmath}
\usepackage{amssymb}
\usepackage{hyperref}
\usepackage{graphicx}
\usepackage{authblk}
\usepackage{fancyhdr}
\usepackage{lineno}

\begin{document}

\title{Magnetic signals from oceanic tides: new satellite observations and applications}


\author[1]{Alexander Grayver\thanks{Corresponding Author: agrayver@uni-koeln.de}}
\author[2]{Christopher C. Finlay}
\author[2]{Nils Olsen}
\affil[1]{Institute of Geophysics and Meteorology, University of Cologne, Germany}
\affil[2]{DTU Space, Technical University of Denmark, Denmark}





\maketitle

\begin{abstract}
Tidal flow of seawater across the Earth's magnetic field induces electric currents and magnetic fields within the ocean and solid Earth. The amplitude and phase of the induced fields depends on electrical properties of both the seawater and the solid Earth, thus can be used as a proxy to study seabed properties or potentially for monitoring long-term trends in the global ocean climatology. This paper presents new global oceanic tidal magnetic field models and their uncertainties for four tidal constituents, including $M_2, N_2, O_1$ and $Q_1$, which was not reliably retrieved previously. Models are obtained through a robust least-squares analysis of magnetic field observations from the \textit{Swarm} and CHAMP satellites using a specially designed data selection scheme. We compare the retrieved magnetic signals with several alternative models reported in the literature. Additionally, we validate them using a series of high-resolution global 3-D electromagnetic simulations and place constraints on the conductivity of sub-oceanic mantle for all tidal constituents, revealing an excellent agreement between all tidal constituents and the oceanic upper mantle structure.
\end{abstract}



\section{Introduction}

Flow of an electrically conductive fluid across a magnetic field results in an electric current and electromotive forces \cite{faraday1832vi}. This electrodynamic phenomenon occurs in different parts of the Earth system including the outer core, ionosphere and the oceans. In the case of the oceans and ionosphere, a major forcing mechanism that drives the fluid (seawater or ionized gas, respectively) are gravitational tides arising from the relative motions and gravitational forces in the Earth-Moon-Sun system \cite{darwin1891v}. 


The tidal forcing contributes energy at specific frequencies, referred to as tidal harmonic constituents (for brevity, we will use this term for all spectral lines where tidal gravitational forcing inputs energy), which can be accurately predicted from the development of the tide-generating potential \cite{doodson1921harmonic, cartwright1971new}. Tidal constituents are most often distinguished based on their frequency (e.g. quarter-diurnal, semi-diurnal, diurnal and long-period) and/or origin (e.g. lunar, solar or nonlinear/mixed tides). The strongest tidal constituents and some of their characteristics are listed in Table~\ref{tab:tides}. Note that effects associated with the bathymetry, ocean stratification and other hydrodynamic effects result in the redistribution of the energy and local variations in height and velocity, but they do not alter harmonic frequencies, making tides highly predictable. A more refined and complete description of oceanic tides and their properties can be found in the dedicated literature \cite{egbert2017tidal, stammer2014accuracy}. 

\begin{table}[]
\centering
\caption{Some diurnal and semi-diurnal tidal constituents and their origins. Reworked from \cite{platzman1971ocean}.}
\begin{tabular}{|c|c|c|c|c|}
\hline Symbol & Period (h) & Derived from$^\dagger$ & Origin & $C^\ddag$ \\
\hline \multicolumn{5}{|c|}{ Semidiurnal tides } \\
\hline $\mathrm{K}_2^{\mathrm{L}}$ & 11.967 & $2 \omega_L+2 \omega_1(=2 \Omega)$ & declinational to $\mathrm{M}_2$ &  0.0768 \\
\hline $\mathrm{K}_2^{\mathrm{S}}$ & 11.967 & $2 \omega_{\mathrm{S}}+2 \omega_2(=2 \Omega)$ & declinational to $S_2$ & 0.0365 \\
\hline $\mathrm{S}_2$ & 12.000 & $2 \omega_S$ & principal solar & 0.4299 \\
\hline $\mathrm{M}_2$ & 12.421 & $2 \omega_L$ & principal lunar & 0.9081 \\
\hline $\mathrm{N}_2$ & 12.658 & $2 \omega_L-\left(\omega_1-\omega_3\right)$ & elliptical to $\mathrm{M}_2$ & 0.1739 \\
\hline $\mathrm{L}_2$ & 12.192 & $2 \omega_L+\left(\omega_1-\omega_3\right)$ & elliptical to $\mathrm{M}_2$ & 0.0257 \\
\hline \multicolumn{5}{|c|}{ Diurnal tides } \\
\hline $\mathrm{K}_1^{\mathrm{L}}$ & 23.934 & $\left(\omega_L-\omega_1\right)+2 \omega_1(=\Omega)$ & declinational to $\mathrm{O}_1$ & 0.3623 \\
\hline $\mathrm{K}_1^{\mathrm{S}}$ & 23.934 & $\left(\omega_{\mathrm{S}}-\omega_2\right)+2 \omega_2(=\Omega)$ & declinational to $P_1$ &  0.1682 \\
\hline $\mathrm{P}_1$ & 24.066 & $\left(\omega_{\mathrm{S}}-\omega_2\right)$ & principal solar &  0.1755 \\
\hline $\mathrm{O}_1$ & 25.819 & $\left(\omega_L-\omega_1\right)$ & principal lunar & 0.3769 \\
\hline $\mathrm{Q}_1$ & 26.868 & $\left(\omega_L-\omega_1\right)-\left(\omega_1-\omega_3\right)$ & elliptical to $\mathrm{O}_1$  & 0.0722 \\
\hline
\end{tabular}
\newline
$^\dagger$ Frequency notation is derived from the fundamental astronomical periods defined as: $2\pi\Omega^{-1} = 23.9344~\mathrm{h}$ -- sidereal day; $2\pi\omega_s^{-1} = 24.0~\mathrm{h}$ -- mean solar day; $2\pi\omega_L^{-1} = 24.8412 \;\mathrm{h}$ -- mean lunar day; $2\pi\omega_1^{-1} = 27.3216 \;\mathrm{days}$ -- period of lunar declination; $2\pi\omega_2^{-1} = 365.2422 \;\mathrm{days}$ -- period of solar declination; $2\pi\omega_3^{-1} = 8.847 \;\mathrm{years}$ -- period of lunar perigee.
\newline
$^\ddag$ The coefficient, $C$, gives a global measure of each constituent's relative portion of the tide-generating potential.
\label{tab:tides}
\end{table}

The atmosphere (including the ionosphere) is also subject to gravitational tides. However, in addition to the gravity forces, the ionosphere experiences daily thermal tides generated by solar heating. These convective motions of the ionosphere dominate the motions generated by gravitational forces. As a result, motionally induced electric currents due to gravitational tides are dwarfed by the currents generated by thermal winds. Additionally, the electrical conductivity of the ocean remains virtually constant on diurnal time scales, with the exception of the Sea Surface Temperature and Salinity \cite{kawai2007,fine2015}, although these near-surface variations have a negligible effect on tidal magnetic signals that are primarily controlled by the depth-integrated seawater conductance \cite{trossman2019}. In contrast, the conductivity structure of the ionosphere varies strongly with the local time and depends on upstream solar wind conditions \cite{yamazaki2017sq}. In addition, the ionospheric conductivity is highly anistropic. These aspects are important when considering the design of optimal data selection schemes and the validation of results, particularly for tidal constituents with periods very close or equal to 24 and 12~hours.  

The phenomenon of electric currents induced by moving fluids, hereinafter referred to as \textit{motional induction}, was predicted by Faraday shortly after the formulation of the law that bears his name \cite{faraday1832vi}. First successful observations of this phenomenon were reported already in the 19th century, based on measurements of the electric potential along telegraph cables or wires grounded near or in the ocean. These early observations established that the measured potential contains both irregular and periodic components with the latter varying at solar and lunar (semi-)diurnal periods. For a review of these early works, the reader is referred to \cite{longuet1949electrical}. 

Early works in terrestrial magnetism also recognized solar and lunar components in the daily variations of the geomagnetic field (see a discussion in \cite{chapman1919solar}). These variations were mostly ascribed to the ionospheric dynamo, and a smaller internal component of the observed diurnal variations was explained by electromagnetic induction effects in the solid Earth \cite{schuster1889}. A comprehensive review of earlier works dedicated to studying the geomagnetic tides of external origin is presented in \cite{winch1981spherical}. The presence of the intrinsic ocean-generated component in observed geomagnetic tides was discussed by W. van Bemmelen in the work titled "Die lunare Variation des Erdmagnetismus" published in 1912 (the original study \cite{vanBemmelen1912} appears not to be available online, a brief summary can be retrieved from the \textit{in memoriam} \cite{vanBemmelen1943}). Decades later, a systematic analysis of coastal, inland and the first ocean-bottom observations \cite{larsencox66, filloux1967ocean, malin1970separation, chave1989observations}, as well as development of an analytic theory describing the responsible variations \cite{longuet1954electrical, larsen1968electric, Sanford71} allowed many groups to model oceanic tidal magnetic signals and validate the growing body of observations. In the majority of early geomagnetic field studies, ocean-induced magnetic fields were seen as noise or a nuisance component rather than the signal of interest \cite{malin1970separation}. Therefore, little attention was paid to this part of the observed signal beyond proving its existence. Nevertheless, some early works did recognize the potential of using oceanic tidal magnetic fields for studying the electric properties of the seabed, upper mantle or even seawater currents \cite{longuet1949electrical, larsen1968electric, tyler1997geophysical}.

A major upsurge in the interest of the geophysical community came with the satellite era. The magnetic signature of a principal lunar $M_2$ tide was isolated in satellite magnetic field observations a few years into the satellite mission CHAMP \cite{tyler2003satellite}. Later, a CHAMP-based comprehensive geomagnetic field model CM5 \cite{sabaka2015cm5} used a potential field representation of the $M_2$ tidal magnetic signal and delivered a global model of the $M_2$ tide magnetic signal (data from earlier missions SAC-C and {\O}rsted were also included, although they play a less significant role in characterizing oceanic magnetic signals due to a weaker signal at their higher altitudes). This model was accurate enough to constrain the global electrical conductivity of the sub-oceanic upper mantle for the first time \cite{schnepf2015can, grayver2016satellite}. The science program of the ESA \textit{Swarm} mission, launched in November 2013, defined magnetic signals due to oceanic tides  as a mission science objective \cite{olsen2013swarm} and magnetic signals due to the dominant $M_2$ tide have indeed been retrieved from the  \textit{Swarm} satellite data \cite{sabaka2016extracting, sabaka2018comprehensive}. The work of \cite{grayver2019magnetic} combined both CHAMP and \textit{Swarm} data and, in addition to the $M_2$ constituent, also added $N_2$ and $O_1$ tides. Subsequent works by other groups also succeeded in extracting $M_2, N_2$, and $O_1$ components \cite{sabaka2020cm6, saynisch2021tide}. All these studies adopted a potential field representation for the tidal magnetic signals as in \cite{sabaka2015cm5} and used vector, scalar and approximate gradients (along-track differences for the CHAMP and both along-track as well as East-West differences between Alpha and Charlie satellites for the \textit{Swarm} constellation) data. On the other hand, these studies differ in the details of the data selection and geomagnetic field modelling approaches. In short, our previous work \cite{grayver2019magnetic} used the so-called sequential approach, where the magnetic field residuals used for the global tidal harmonic analysis were obtained by subtracting models of core and magnetospheric fields as given by the CHAOS-6 model \cite{finlay2016recent}. This approach will be further developed in the present study. In contrast, the works of \cite{sabaka2020cm6} and \cite{saynisch2021tide} used a Comprehensive Inversion scheme handling all sources simultaneously and a method based on a Kalman filter, respectively. 

In this study we will address the problem of mapping the magnetic signals due to oceanic tides with satellite observations, characterize the uncertainty of the obtained models and present systematic comparisons with detailed 3-D electromagnetic simulations and other models reported in the literature. Finally, we will use new tidal magnetic field models to infer constraints on the electrical conductivity of the lithosphere-asthenosphere system in the sub-oceanic mantle.

\section{Methods}

\subsection{Governing Equations}

Electric currents and magnetic fields induced by oceanic tides occur at frequencies where displacement currents are negligible \cite{larsson2007electromagnetics}. Therefore, generation and evolution of tidal electromagnetic fields can be accurately described by adopting a magneto-quasi static approximation to Maxwell's equations \cite{larsson2007electromagnetics} and assuming a linear non-polarized medium without additional magnetization currents:
\begin{align}\label{eq:maxwell}
\begin{split}
\mu_0^{-1} \nabla \times \vec{B} &= \vec{J},  \\
\nabla \times \vec{E} &= -\textrm{i} \omega \vec{B}, 
\end{split}
\end{align}
where $\vec E$ is the electric field $[\textnormal{V/m}]$, $\vec B$ is the magnetic flux density $[\textnormal{T}]$, $\omega$ is angular frequency, and $\mu_0$ is magnetic permeability of the vacuum $[\textnormal{Vs/Am}]$. Our modelling domain is a sphere with a 3-D distribution of the electrical conductivity $\sigma \equiv \sigma(\vec{r}) \,\, [\textnormal{S/m}]$. $\vec{r}=(r, \theta, \phi)$ denotes a position vector in the Earth centered spherical coordinate system with $r$, $\theta$ and $\phi$ being distance from the centre, polar and azimuthal angles, respectively. We assume that fields decay to zero sufficiently fast as $r \rightarrow \infty$ and satisfy the Helmholtz theorem \cite{griffiths2017introduction}. The dependence of the variables on spatial coordinates is implied. The current density term, $\vec{J} \; [\textnormal{A/m}^2]$, is specified using Ohm's law in a moving conductor \cite{jackson1999classical}
\begin{align}
\begin{split}
\label{eq:motional_current}
\vec{J} &= \sigma\left[\vec{E} + \vec{v} \times \vec{B}\right]\\
 &\approx \sigma\left[\vec{E} + \vec{v} \times \vec{B}_c\right]
\end{split}
\end{align}
where $\vec{v} \; [\textnormal{m/s}]$ is the fluid velocity ($\vec{v} = 0$ outside of the ocean in our case). The second line in the eq. (\ref{eq:motional_current}) follows by observing that the ocean-induced magnetic field, $\vec{B}_o \; (\lessapprox 5-10~\mathrm{nT})$, is much smaller than the Earth's main magnetic field, $\vec{B}_c \; (\approx 30,000-60,000~\mathrm{nT})$, mostly originating in the outer core. This definition makes the current generating term an extraneous source and linearizes the system relative to the source.

Since tides are periodic phenomena, all the equations above were stated in frequency domain assuming that fields, currents and the velocity vary with angular frequency $\omega$ and adopting the $\textnormal{exp}(\mathrm{i}\omega t)$ time dependency. 


\subsection{3-D Numerical Simulations}
\label{sec:numsim}

A direct numerical simulation (DNS), hereinafter denoted by $\vec{B}^{\mathrm{DNS}}(\vec{r})$, can be obtained by solving eqs. \ref{eq:maxwell}-\ref{eq:motional_current} numerically. This section provides details on how DNS is performed in this study and when it is employed.

In subsequent sections, we will use DNS of ocean tidal induced magnetic fields from different tidal constituents for two purposes: (i) to validate the observed signals and (ii) to test sensitivity of the observed signals to the conductivity contrast across the lithosphere-asthenosphere boundary. Since tidal magnetic signals are strongly affected by lateral electrical conductivity contrasts \cite{velimsky2018}, a realistic 3-D model of the ocean and marine sediments conductivity is required. Accordingly, a full 3-D EM induction solver is needed to account for all inductive and galvanic effects for a complex tidal electrical current density and a 3-D electrical conductivity distribution. In this study, we rely on a novel 3-D EM code \cite{grayver2019three} that is based on the adaptive Finite Element Method implemented using the deal.II library \cite{arndt2020deal} and solves eqs. (\ref{eq:maxwell}) numerically in a sphere using an optimal multi-grid solver \cite{grayver2015large}. A hexahedral locally refined mesh was used and the solution is approximated by second-order $curl$-confirming N{\'e}d{\'e}lec basis functions \cite{schoberl2005high} within each grid cell. This results in uniform horizontal resolution of $\approx 10$~km at the surface and at satellite altitude. This is the reference solution for task (i) listed above. Given the limited computational resources, a lower resolution grid was used for the sensitivity study (ii) where many forward solutions are required.

For the electrical conductivity of the ocean and marine sediments, we used the realistic 3-D model of \cite{grayver2021global}, which relies on validated observations of ocean climatology from the World Ocean Atlas and the thermodynamic equation of state for seawater (TEOS-10). In future, this can be updated with a validated product \cite{tyler2017electrical, reagan2019world}. For land areas, the model of \cite{alekseev2015compilation} was taken. The underlying mantle section is assigned a new radial reference conductivity profile derived from the joint inversion of recent \textit{Swarm}-based magnetospheric and $M_2$ tidal magnetic signals (specifically, Comprehensive Inversion \textit{Swarm} products \textit{MMA\_0906} and \textit{MTI\_0901}) following the methodology described in our earlier work \cite{grayver2017joint}. The TPXO-9 model was used for the velocities of the tidal constituents \cite{egbert2002efficient}. The magnetic field $\vec{B}_c$ was calculated using the IGRF-11 model evaluated at 1st November 2014. 

\subsection{Observational Models}
\label{sec:modelparams}

Assuming that the magnetic field is curl-free at a satellite location, we can represent the ocean-induced magnetic field through a scalar potential as $\vec{B}^{\mathrm{obs}} = -\nabla V$ \cite{gauss1877}. For a given location $\vec{r}$ and at a time $\Delta t$ taken with respect to the Greenwich phase, the scalar potential is given by
\begin{equation}
\label{eq:potential_exp}
V(\Delta t, \vec{r}) = \mathcal{R}\left\{\exp{(i\omega \Delta t)} R_E \sum_{n = 1}^{N_\textnormal{max}} \left( \frac{R_E}{r} \right)^{n+1} \sum_{m = -n}^n \tau_n^m P_n^{|m|}(\cos{\theta})\exp{(i m\phi)}\right\},
\end{equation}
where $R_E = 6371.2$~km is the Earth's reference radius; $\tau_n^m$ are complex spherical harmonic (SH) coefficients of degree $n$ and order $m$; $P_n^{|m|}$ are Schmidt semi-normalized associated Legendre functions; and $\mathcal{R}\left\{\cdot\right\}$ denotes the real part. While complex-valued notation is more compact to state, our implementation uses a real-valued equivalent as elaborated previously in \cite{sabaka2018comprehensive, grayver2019magnetic}.

By calculating frequencies $\omega$ from the periods given in Table~\ref{tab:tides}, SH coefficients for each harmonic tidal constituent were estimated up to a maximum spherical harmonic degree $N_\textnormal{max}$ from pre-processed satellite geomagnetic field observations (see Section \ref{sec:data}). The values for $N_\textnormal{max}$ were chosen experimentally for each constituent to ensure reasonable signal-to-noise levels while retaining the highest possible spatial resolution. A Huber-weighted robust least-squares method was used to retrieve the SH coefficients $\tau$ (the method is documented in our previous study \cite{grayver2019magnetic}). Hereinafter, the radial magnetic field due to a given tidal constituent at a location $\vec{r}$ above the ground, synthesized from corresponding estimated SH coefficients using eq. (\ref{eq:potential_exp}), will be denoted as $B_r^{\mathrm{obs}}(\vec{r})$.

To quantify errors of a SH model given by coefficients $\tau$, we perform a simple error propagation. Assuming that $\mathbf{A}$ is a matrix relating SH coefficients to the observations for a given tidal constituent and $\mathbf{W}$ is a diagonal matrix containing Huber weights, the model covariance matrix $\mathbf{C}_{\tau} = \left(\gamma^{-2}\mathbf{A}^T\mathbf{W}\mathbf{A}\right)^{-1}$ was calculated for each tidal constituent, with $\gamma = 2$~nT being the assumed prior data standard deviation, consistent with CHAOS-7 residual statistics (cf. Figure~6 in \cite{finlay2020chaos}) for CHAMP and \textit{Swarm} vector field components. $\delta\mathbf{\tau}=\mathrm{diag}(C_{\tau})^{1/2}$ are then the model standard deviations, which we use as a measure of the model uncertainty propagating it to magnetic fields. Since Huber weights and periods are different for different constituents, this procedure is repeated for each tidal constituent independently.

\section{Data}
\label{sec:data}

In our previous work \cite{grayver2019magnetic} we took a simple approach and  adopted the same data selection scheme as used to construct the CHAOS-6 geomagnetic field model \cite{finlay2016recent}. In the present study we have implemented an improved data selection scheme specifically designed for extracting the oceanic tidal signal. 

We make use of both vector magnetic field and scalar field intensity measurements from the CHAMP satellite (L3 data version) between 28th July 2000 and 3rd September 2010, and the three \textit{Swarm} satellites Alpha, Bravo and Charlie (data product MAGx\textunderscore LR\textunderscore 1B, version 0602) between 26th November 2013 and 31st October 2022.

In addition we use gradients of the scalar and vector fields from CHAMP and \textit{Swarm}, approximated in the along-track direction by differences between data separated by 15 seconds, and approximated in the east-west direction by differences between data from \textit{Swarm} Alpha and corresponding data from \textit{Swarm} Charlie that is closest in latitude, collected within 50 seconds but typically separated in time by about 10 seconds \cite{Olsen2015SwarmInit}. 


Magnetic field measurements were used with a 15 second sampling cadence from both CHAMP and \textit{Swarm}.  Vector components were transformed from the frame of the vector magnetometer to the north-east-center frame using star tracker attitude data and in-flight corrections estimated using the CHAOS-7.12 model.

\begin{figure}[h]
\includegraphics[width=\textwidth]{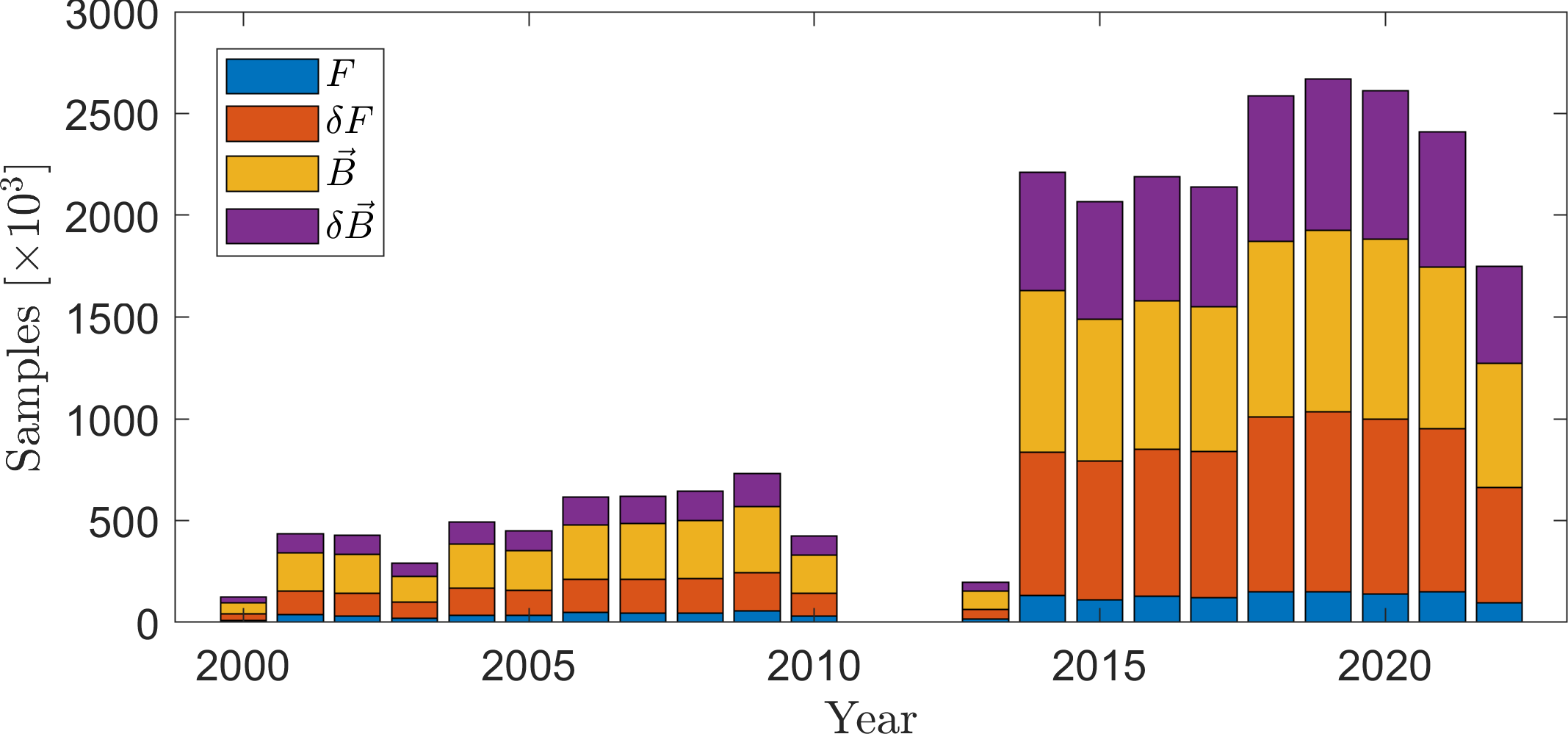}
\caption{Number of observations used in the analysis as a function of time. Scalar and vector observations are denoted by $F$ and $\vec{B}$, respectively. The corresponding scalar and vector magnetic field gradients are denoted $\delta F$ and $\delta \vec{B}$. Each component of a vector field counts as a single datum.}
\label{fig:datahist}
\end{figure}

 Vector field data and gradients were used below 55~degrees quasi-dipole latitude \cite{laundal2017magnetic}; at higher quasi-dipole latitudes (i.e. polar regions) only scalar field intensity data and gradients were used. Vector field data and gradients were excluded from polar regions in order to avoid contamination by field aligned currents. At high latitudes field aligned currents primarily affect the vector field components perpendicular to field lines and not the scalar field intensity \cite{langel1974,langel1980}. Only data from geomagnetically quiet times were used.  Specifically, data were excluded when the disturbance index $Kp\,>\,3o$ or ${\frac{d}{dt}}\,|RC(t)|\,>\,3\,$nT/hr, where $RC(t)$ is an index monitoring the strength of the magnetospheric ring current (RC) \cite{Olsen:2014}. For scalar data, vector data and vector gradients, we used data only from dark regions (sun at least 10 degrees below the horizon). We used scalar gradients from both dark and sunlit regions, except within 10 degrees of the magnetic equator for sunlit regions \cite{Olsen2015SwarmInit,olsen2017lcs1}. The final distribution of the employed satellite observations as a function of time is presented in Figure~\ref{fig:datahist}.

In order to focus on the oceanic tidal signal, the time-dependent internal (core) field and magnetospheric fields from the CHAOS-7.12 model \cite{finlay2020chaos}, and the lithospheric field from the LSC-1 field model \cite{olsen2017lcs1} were removed from the observations; only the residuals after these corrections were used for constructing the field models described below.  

Compared with the CHAOS-6 model used in our previous study \cite{grayver2019magnetic}, the CHAOS-7 model employed during the data processing here involves a number of improvements.  The basic field model parameterization is similar but it was constructed using updated datasets from the three \textit{Swarm} satellites and ground observatories, as well as a new source of vector magnetic field data from the Cryosat-2 satellite. CHAOS-7 also uses stricter geomagnetic quiet time data selection criteria compared with CHAOS-6, relaxed temporal regularization of the high degree internal field and uses the diagonal part a 3D Q-response matrix \cite{grayver2021time} when induced fields are computed.  Full details of the CHAOS-7 model can be found in the reference publication \cite{finlay2020chaos}; CHAOS-7.12 was the most up to data version of the model available at the time of data processing, it was 12th update of the CHAOS-7 model made using the \textit{Swarm} and ground observatory data as available in September 2022.

\section{Results}

\subsection{Tidal magnetic signals}

Tidal magnetic signals are harmonic oscillations, thus the magnetic fields calculated from complex-valued SH coefficients $\tau_X$ for a given tidal constituent $X \in \{M_2, N_2, O_1, Q_1\}$ can be presented either by using the real and imaginary parts or their amplitude and phase angle on a regular global grid at a fixed radius. We chose to work with the real/imaginary representation and, following the established practice, used the radial magnetic field component for all comparisons and figures below. Whenever we refer to the field or SH spectrum at the satellite altitude, the constant value of 430~km above the Earth's reference radius $R_E$ was used, representing the approximate mean altitude of the spacecraft. Numerical solutions used for comparison and validation were calculated for each tidal constituent following the description in the Methods section. 

We will compare our models to numerical simulations and results from other studies. For brevity, the following acronyms for existing models will be used:
\begin{itemize}
\item GO19: our previous models \cite{grayver2019magnetic}.
\item CM6: comprehensive model version six \cite{sabaka2020cm6}.\item CI9: comprehensive inversion product version MTI\_0901 for the $M_2$ tide, based on  9~years of \textit{Swarm} data and following the methodology of \cite{sabaka2018comprehensive}.
\item KALMAG: model of \cite{julien2022kalmag}, also presented in \cite{saynisch2021tide}. Provided SH coefficients were recalculated with the updated data set including \textit{Swarm} data up to November 2022 (Dr.~J.~Baerenzung, pers.\ comm.).
\item GFO24 and GFO24-HR: this study, two sets of models with lower and higher $N_{\mathrm{max}}$ values were derived (see Table~\ref{tab:rmsd}). 
\end{itemize}
Not all models used the same data. In particular, GO19 and CM6 did not include  \textit{Swarm} observations from the last solar minimum, in particular the years 2019-2021. Given the similarities between the CI and CM models, we consider the more recent CI results, CI9, to be representative of what one can achieve with longer \textit{Swarm} time series. However, CI9 provides only a model for $M_2$, whereas CM6 includes $M_2, N_2$ and $O_1$. Currently, the signal due to the $Q_1$ tide is provided only by the GFO24 and KALMAG models. The highest available SH degree was used to calculate observed signals from models, except for  CM6, where lower values (right Table~\ref{tab:rmsd}) were taken due to suspected noise contamination at the highest degrees as reported in the original study \cite{sabaka2020cm6}.

\begin{figure}[h]
\includegraphics[width=\textwidth]{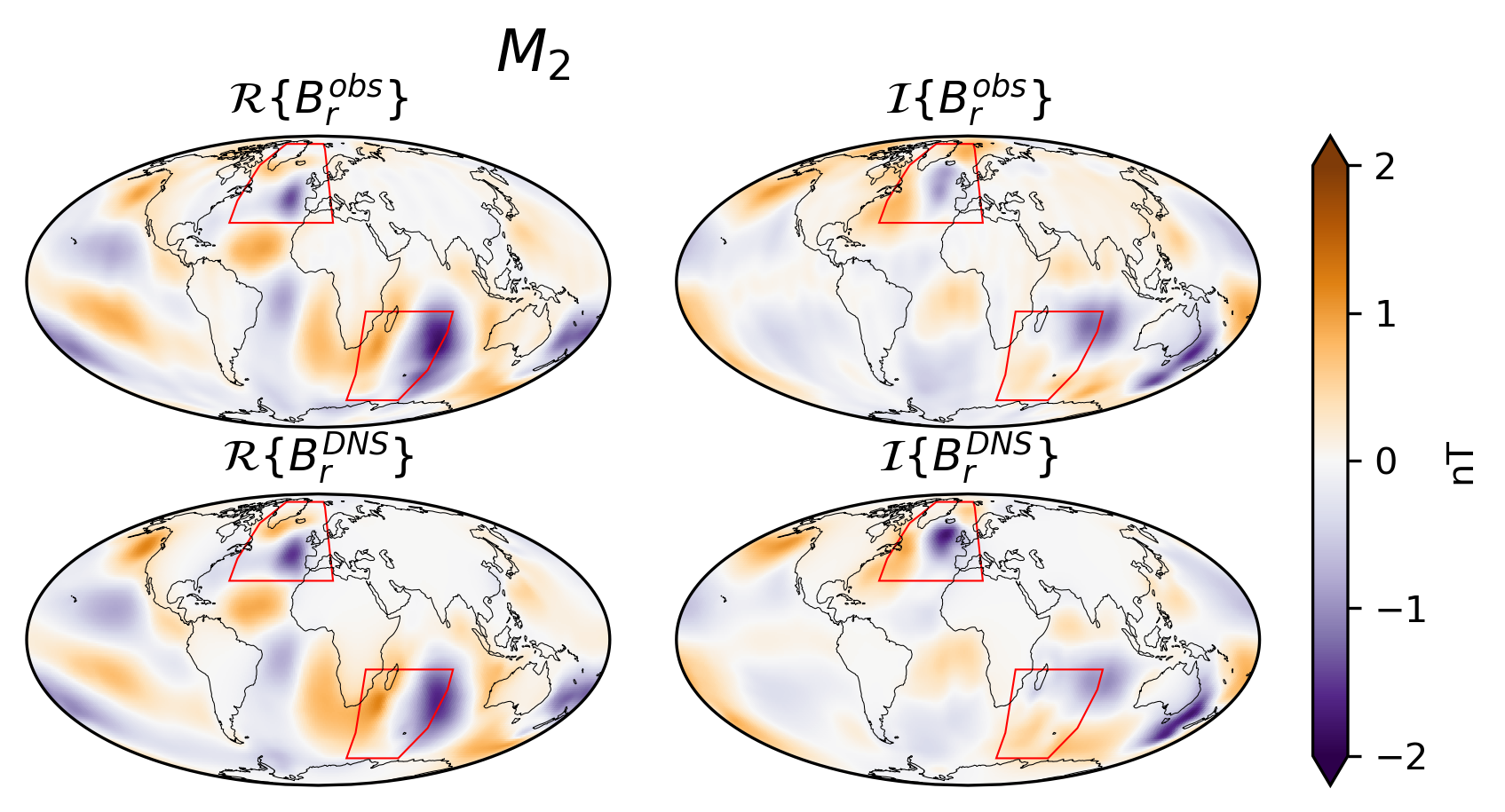}
\includegraphics[width=\textwidth]{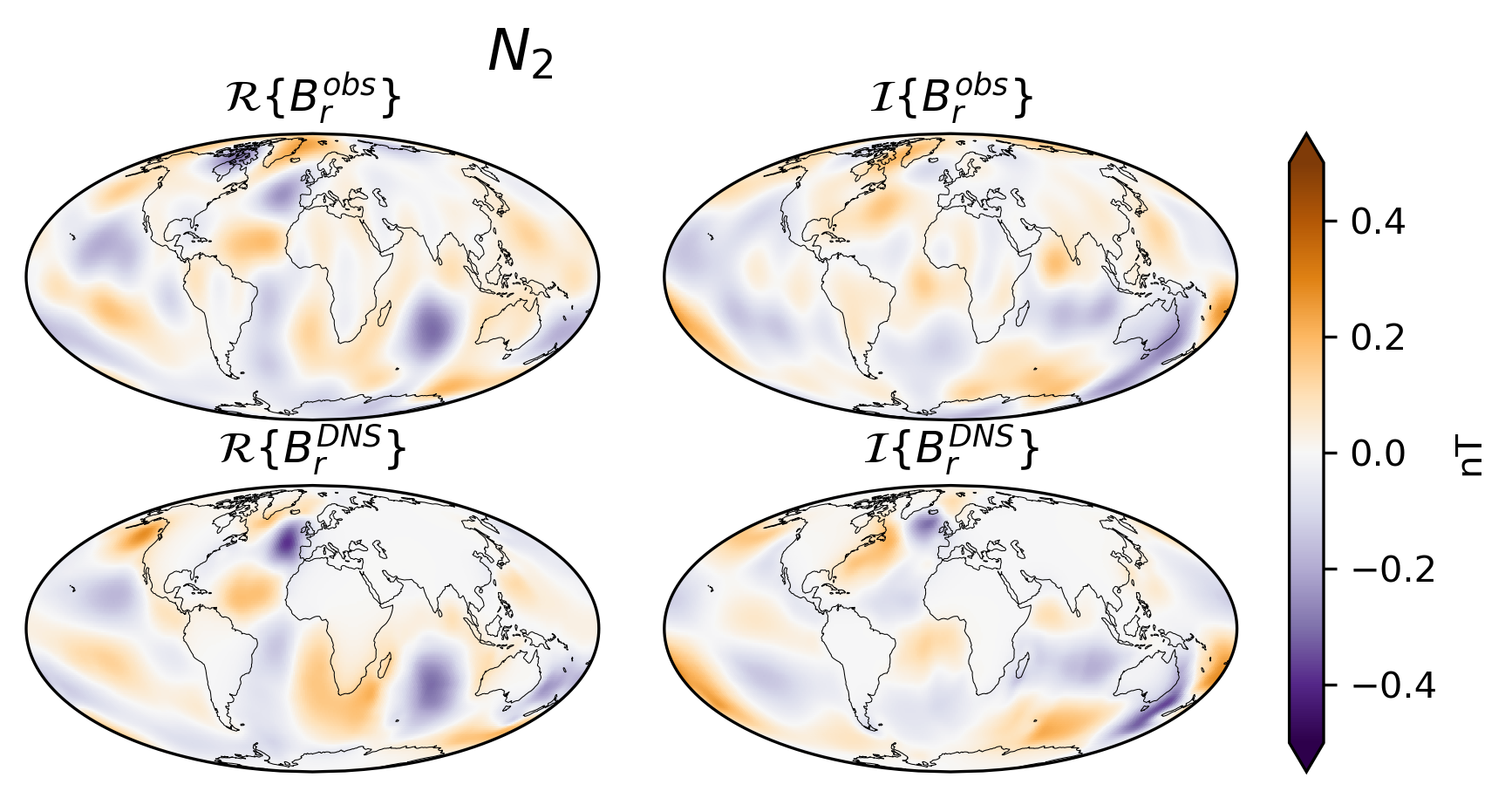}
\caption{Maps of the real and imaginary parts of the observed, $B_r^{\mathrm{obs}}$, as given by the GFO24-HR model, and numerically simulated, $B_r^{\mathrm{DNS}}$, ocean tidal radial magnetic fields due to the semi-diurnal $M_2$ and $N_2$ tidal constituents. Fields are given at the altitude of 430~km. Red rectangles highlight the regions discussed in detail in the text.}
\label{fig:Br_semidiurnal}
\end{figure}

\begin{figure}[h]
\includegraphics[width=\textwidth]{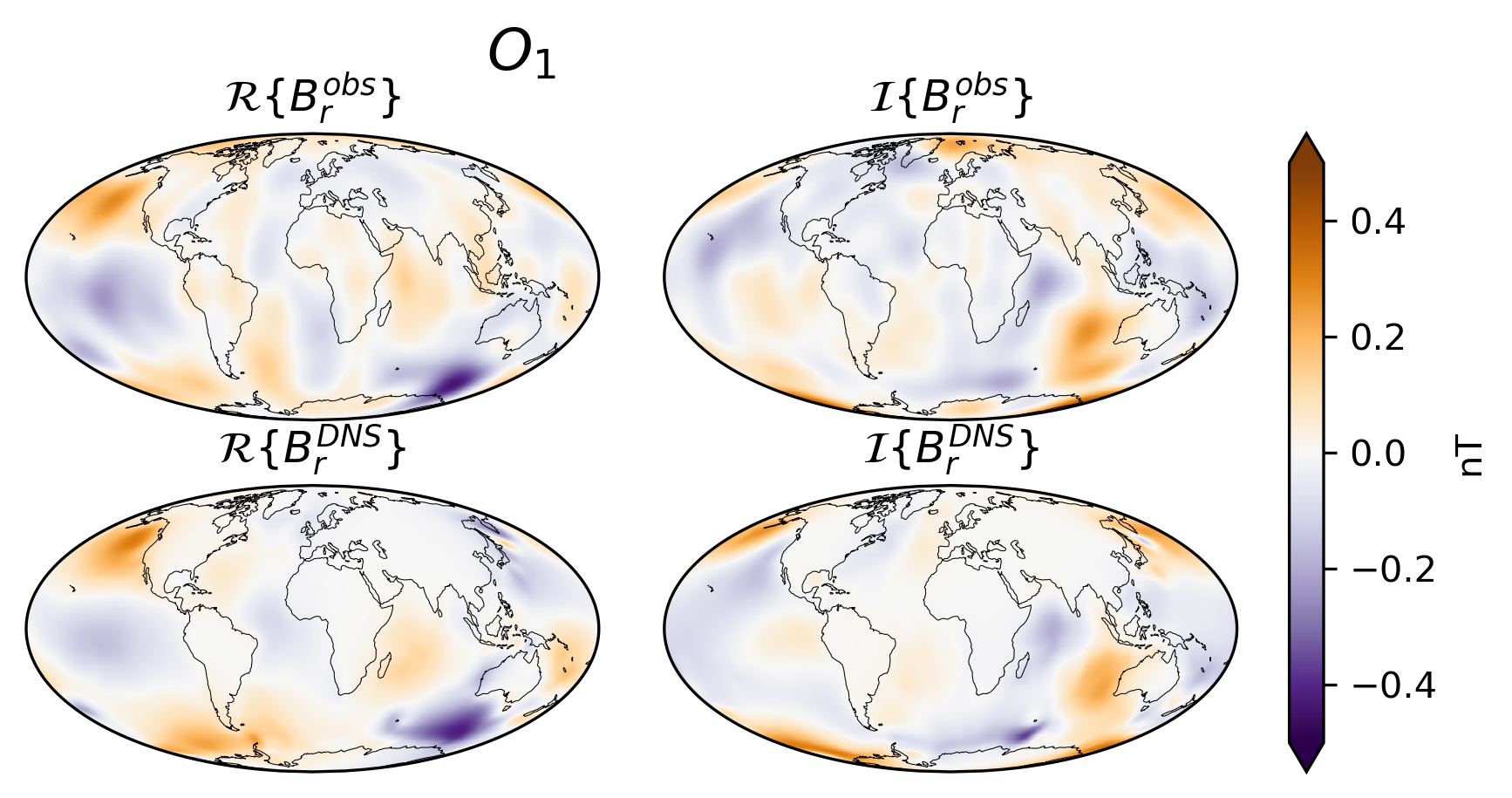}
\includegraphics[width=\textwidth]{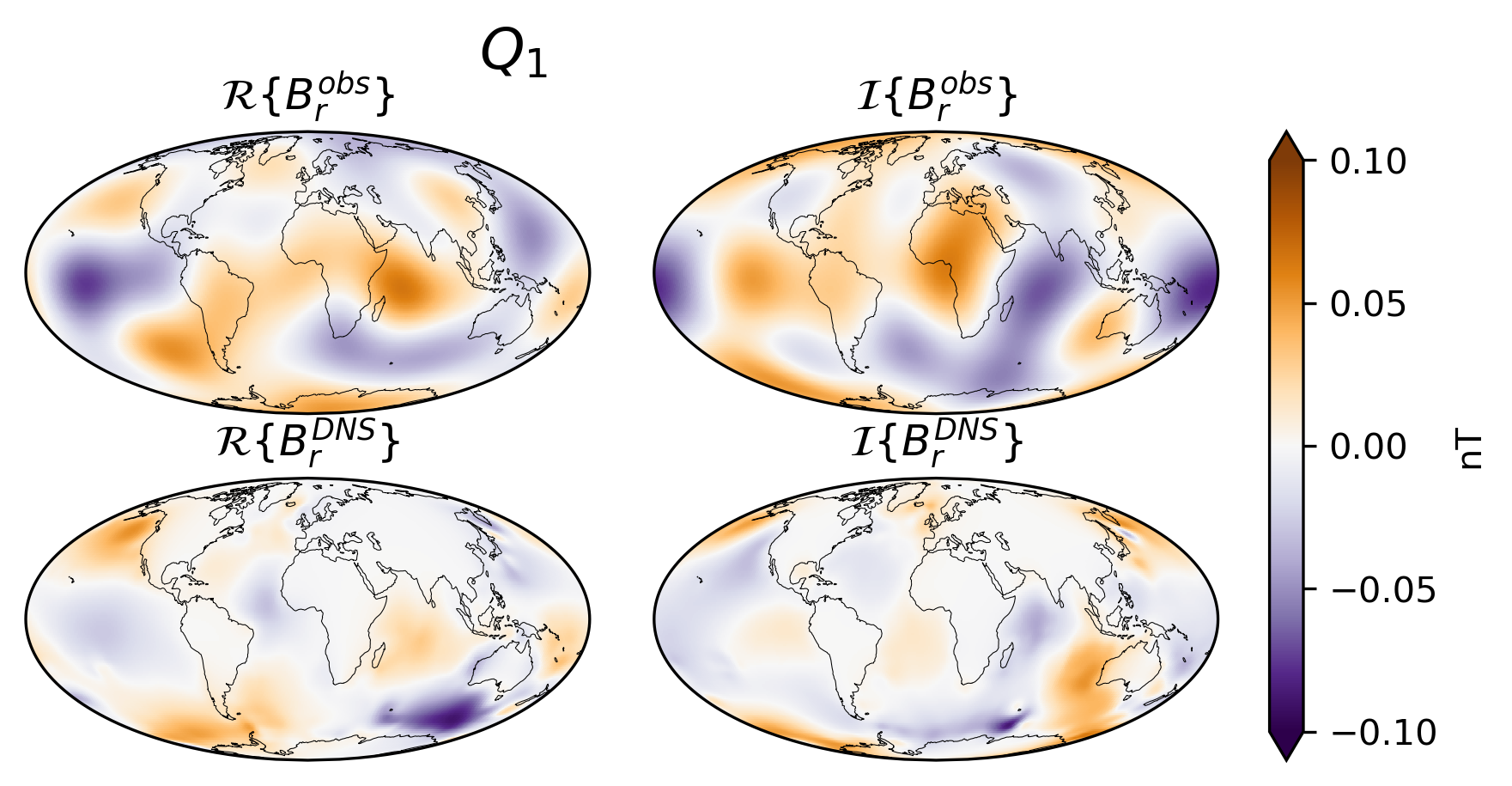}
\caption{Same as Figure~\ref{fig:Br_semidiurnal}, but for the diurnal $O_1$ and $Q_1$ tidal constituents.}
\label{fig:Br_diurnal}
\end{figure}

Figures \ref{fig:Br_semidiurnal}-\ref{fig:Br_diurnal} show maps of the radial magnetic field components, at satellite altitude, derived in this study. Both observed and modelled fields are shown and can be compared. Compared to our previous study, GO19, significant improvements in signals at high latitudes are achieved. For instance, strong tidal signals in the Greenland-Iceland region (most visible in  $M_2$) are now reliably retrieved and exhibit a good agreement with simulations. However, one clear sign of  contamination is significant magnetic signals far from the oceans, deep inland of the continents, mostly caused by ionospheric currents. Contamination at high latitudes is more apparent in the magnetic fields retrieved for the $N_2$ and $O_1$ tides, while the agreement at mid and equatorial latitudes is generally excellent. Even the weak $Q_1$ signal exhibits a good degree of similarity between the observed and simulated signals in terms of the general patterns, although the spatial resolution and signal-to-noise ratio remain low. With $N_{\mathrm{max}} = 32$ for the $M_2$ signal our models approach a spatial resolution of $1000-2000$~km at the Earth's surface. Regional features caused by variations in the tidal transport, e.g. around islands, can now be retrieved from data and validated by simulations. Figure~\ref{fig:Br_semidiurnal} highlights a few such regions with strong local gradients in the tidal magnetic fields that are visible in both observations and simulations. For instance a region southeast of Africa with regional amplifications (south of the Madagascar) and weakening of the field (around French Southern and Antarctic Lands) are well characterized.

\begin{figure}[]
\includegraphics[width=\textwidth]{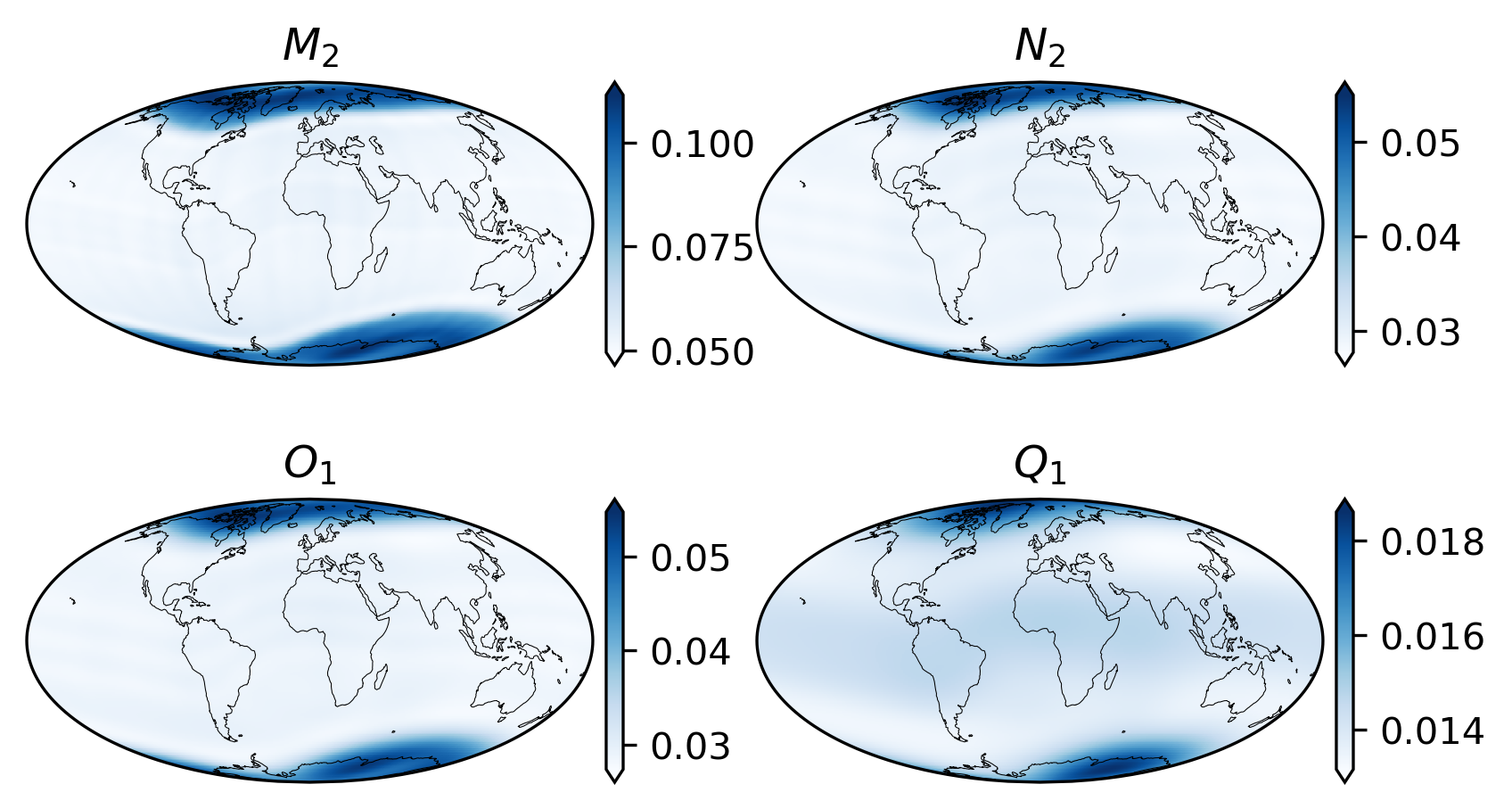}
\caption{Standard deviation (in nT) maps calculated from the model covariance matrix $\mathbf{C}_{\tau}$ for each tidal constituent and propagated to the radial magnetic field at 430 km altitude.}
\label{fig:Br_stdev}
\end{figure}

The corresponding model standard deviations for the radial field are illustrated in Figure~\ref{fig:Br_stdev}. Note that the standard deviation values depend not only on the geometric factors related to the orbit and data distribution, but also on the variance in the original data through the Huber-weights (see Section \ref{sec:modelparams}). Higher values in polar regions result from larger scatter in the observations and because vector field observations in polar regions were excluded. While these formal model standard deviations may be systematically lower than real modelling errors, which also contain correlated noise and potentially systematic biases, the relative variations over globe and between the tidal constituents appear reasonable. Therefore, these standard deviation values can be used in inversion of the tidal magnetic signals for the seabed conductivity (see next Section and studies like \cite{grayver2017joint, sachl2022inversion}). 
To quantitatively compare models with realistic 3-D numerical simulations (see Section \ref{sec:numsim}), we calculate the root-mean-square deviation (RMSD). Assuming that magnetic fields are calculated in the middle of spherical surface elements of size $\Delta\phi \times \Delta \theta$ on a spherical surface at a constant radius, $r$, the RMSD between simulated and observed signals is calculated as
\begin{equation}
\label{eq:rmsd}
\mathrm{RMSD}(r) = \sqrt{\frac{\sum{\left[\delta B_r(\vec{r})\delta B_r(\vec{r})^*\right]\mathrm{d}S}}{\sum{\mathrm{d}S}}},
\end{equation}
where the sum runs over all surface elements (we used a 1$^{\circ}$ resolution, yielding 64,800 surface elements), star denotes the complex conjugation, $\mathrm{d}S = \sin{\theta}\Delta\phi\Delta\theta$ is the surface element area and $\delta B_r(\vec{r}) = B_r^{\mathrm{obs}}(\vec{r}) - B_r^{\mathrm{DNS}}(\vec{r})$. It is necessary to take into account the variable physical area of spherical surface elements to avoid a bias in polar regions.

\begin{table}
\centering
\caption{Left: Root-mean-square deviations (RMSD) between observed tidal signals and 3-D DNS for four considered tidal constituents (calculated as described in Section \ref{sec:numsim}). Radial magnetic field component at the altitude of 430~km was used. Right: corresponding maximum SH degree used to calculate the magnetic fields.}
\label{tab:rmsd}
\begin{tabular}{|l|c|c|c|c|}
\hline
\textbf{RMSD} [nT] & $M_2$ & $N_2$ & $O_1$ & $Q_1$ \\
\hline
GFO24 & 0.166 & 0.063 & 0.070 & 0.027 \\
GFO24-HR$^1$ & 0.167 & 0.069 & 0.074 & 0.033 \\
GO19 & 0.193 & 0.088 & 0.098 & -- \\
CI9 & 0.169 & -- & -- & -- \\
CM6$^2$ & 0.196 & 0.089 & 0.126 & -- \\
KALMAG & 0.210 & 0.088 & 0.092 & 0.062 \\
\hline
\end{tabular}
\begin{tabular}{|l|c|c|c|c|}
\hline
$\mathbf{N}_{\mathrm{max}}$ & $M_2$ & $N_2$ & $O_1$ & $Q_1$ \\
\hline
GFO24 & 28 & 12 & 12 & 4 \\
GFO24-HR$^1$ & 32 & 14 & 14 & 6 \\
GO19 & 28 & 12 & 12 & -- \\
CI9 & 18 & -- & -- & -- \\
CM6$^2$ & 28 & 12 & 12 & -- \\
KALMAG & 30 & 30 & 30 & 30 \\
\hline
\end{tabular}
\newline
$^1$ Higher resolution models, determined up to higher SH degrees compared to GFO24.
\newline
$^2$ Although higher degree expansions were reported in the original study \cite{sabaka2020cm6}, the authors stated that only signals up to the herein specified degrees are reliable.
\end{table}

Table~\ref{tab:rmsd} lists RMSD values for all the compared models. The use of a numerical simulation for the validation has several implications. It is an independent result that relies on the actual equations that simulate all relevant physics. Furthermore, by using water transport from the satellite altimetry-based TPXO-9 model for the electric current term we ensure that simulated motionally induced signals are generated by realistic tidal currents. On the other hand, a key unknown in eqns (\ref{eq:maxwell})-(\ref{eq:motional_current}) is the electrical conductivity of the underlying crust and upper mantle. For this we used a global average radial conductivity profile for the upper mantle derived from the CI-based $M_2$ signal, hence the simulated signals will naturally tend to be consistent with the CI9 model. Although this creates somewhat favourable conditions for the CI9 tidal field model, it is the GFO24 model that exhibits the smallest RMSD. We stress that tidal magnetic signals are also sensitive to lateral conductivity variations in the crust and mantle \cite{sachl2022inversion}, hence non-zero RMSD values can be ascribed to several factors: (i) lateral conductivity variations in the subsurface not accounted for by the present numerical simulation, (ii) errors in the representation of the current generating terms (water transport, core field, seawater conductivity) and (iii) numerical/discretization errors. The last two factors were previously quantified and should not generally exceed $\sim$0.1~nT for the strongest $M_2$ tide \cite{grayver2016satellite, velimsky2018}. 
\begin{figure}[h]
\includegraphics[width=0.45\textwidth]{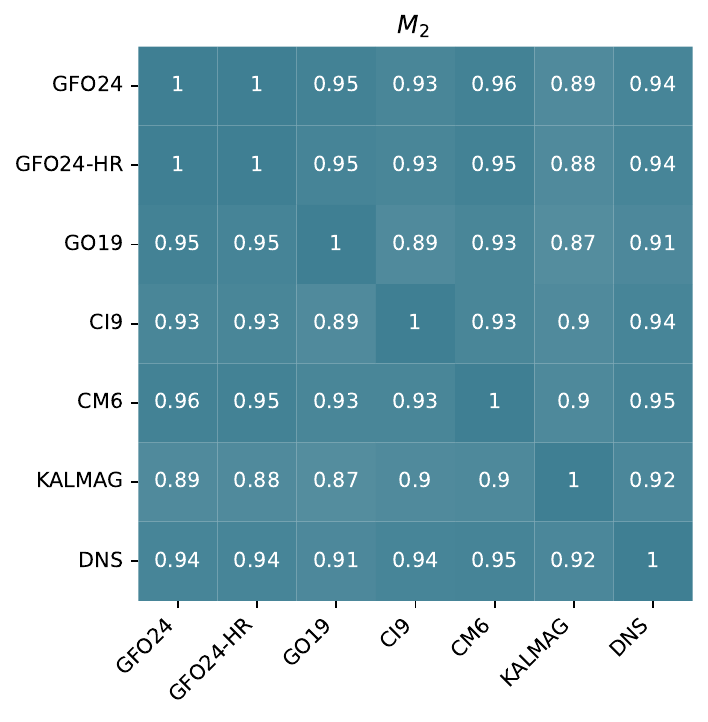}
\includegraphics[width=0.45\textwidth]{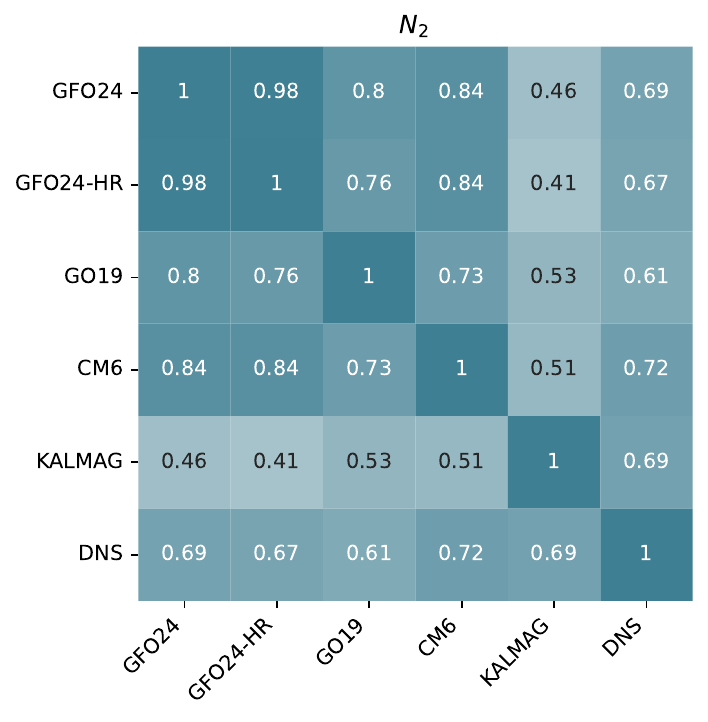}

\includegraphics[width=0.45\textwidth]{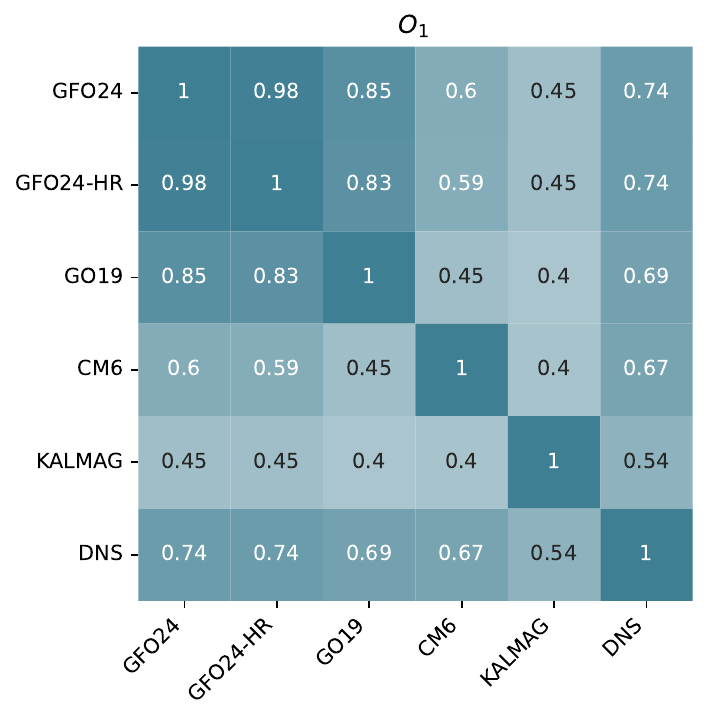}
\includegraphics[width=0.45\textwidth]{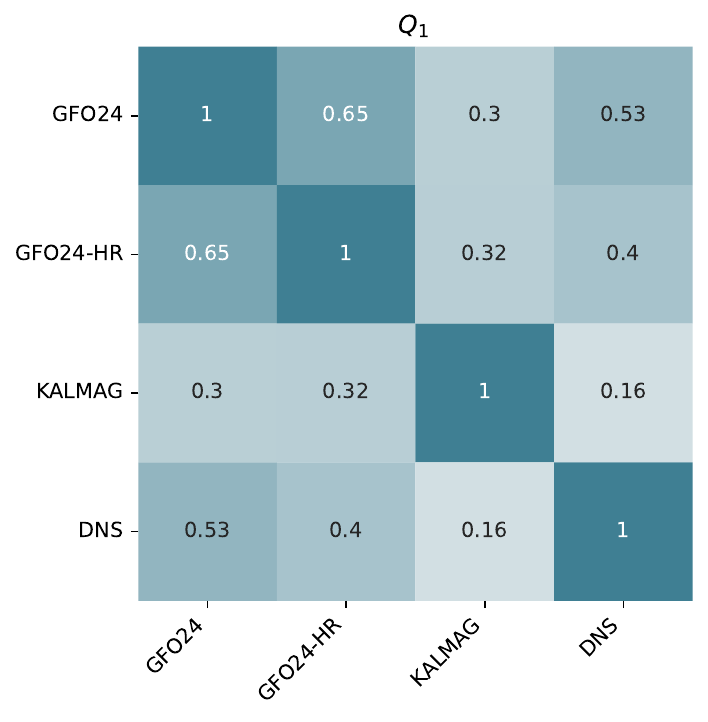}
\caption{Pearson correlation coefficients between different models of the observed signals and a 3-D direct numerical simulation (denoted DNS) for the four tidal constituents. Correlations were calculated between real parts of the radial magnetic field maps at the altitude of 430~km (shown in Figures \ref{fig:Br_semidiurnal}-\ref{fig:Br_diurnal}). Correlation coefficients for imaginary parts are similar.}
\label{fig:xcorr}
\end{figure}
Since estimating absolute values is difficult, their comparison might be less reliable. Therefore, we also looked at cross-correlations. Figure~\ref{fig:xcorr} shows Pearson correlation coefficients between all different models and simulations. The cross-correlation coefficients overall confirm the trends seen in the RMSD values, showing a substantial improvement of the  models constructed in this study over the predecessor study (GO19). Highest correlation coefficients are obtained for the $M_2$ signal. The generally lower correlation coefficients for $Q_1$ are explained by the low resolution and relatively high noise levels, nonetheless the value of $0.53$ between the GFO24 and the simulation is a clear indication that the extracted signal contains the physical signal.

\begin{figure}[h]
\includegraphics[width=\textwidth]{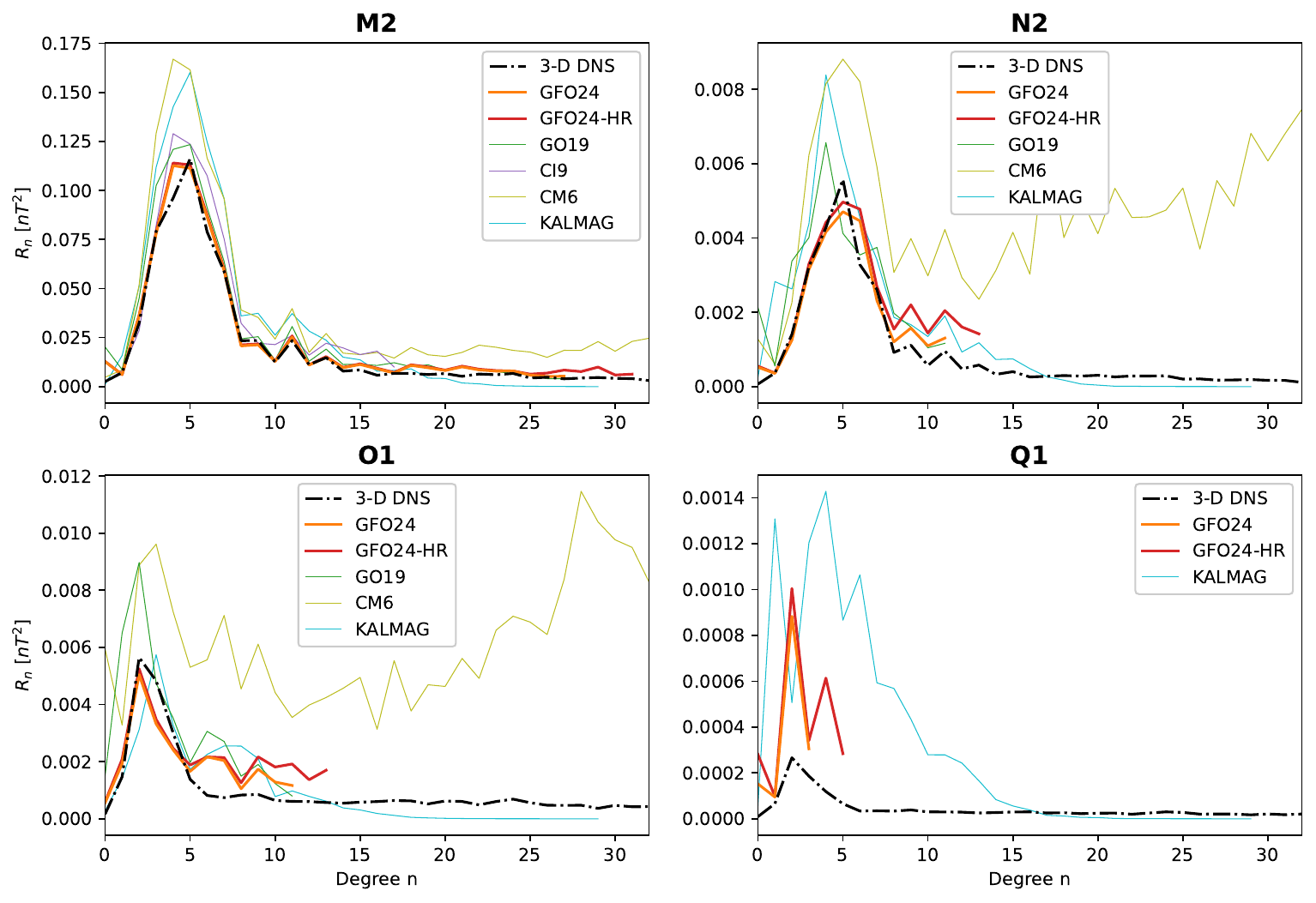}
\caption{The $R_n$ power spectra (eq. \ref{eq:spectra}) of the time-averaged ocean tidal magnetic fields due to four tidal constituents for different models. Black lines denote 3-D numerical simulations (DNS). Spectra were calculated for the reference surface radius $R_E = 6371.2$~km.}\label{fig:spectra}
\end{figure}

Additionally, we compare the models and simulations in terms of their SH power spectra defined at a given radius, $r$, as \cite{sabaka2020cm6}
\begin{equation} 
\label{eq:spectra}
R_n\left( {r}\right) =\left( {n+1}\right) \left( {\frac{R_E}{r}}\right) ^{2n+4}\left\{ {{\frac{1}{2}}\left| {\tau _n^0}\right| ^2+\sum _{m=1}^n\left[ {\left| {\tau _n^m}\right| ^2+\left| {\tau _n^{-m}}\right| ^2}\right] }\right\} .
\end{equation}
Figure~\ref{fig:spectra} shows SH spectra calculated at Earth's surface spherical reference radius for all constituents.

Although the structure of spectra are generally similar, there are some differences: SH degrees $\geq 15-20$ appear to be damped in the KALMAG model, which prevents comparison with other models at these degrees. The new GFO24 models exhibit better agreement with simulations at all scales compared to the predecessor study GO19. In particular, the low degree SH coefficients are closer to the 3-D simulations, highlighting a reduced contamination by external field sources. We stress that the presented SH power spectra were calculated at Earth's surface. Such downward continuation is known to amplify noise \cite{sabaka2010mathematical}, which reveals the intrinsic quality and resolution of the model without the effect of geometric attenuation of higher SH degrees. Strong noise levels, manifested by a growing spectrum with SH degree, or damping due to strong regularization/prior are more apparent when spectra are calculated at the surface. The generally good agreement, even for smaller $N_2$ and $O_1$ constituents, is a remarkable achievement of modern satellite magnetometry. That said, we do observe a larger misfit at higher SH degrees for the GFO24-HR models compared to the GFO24 models where they overlap, indicating higher variances at high SH degrees in the GFO24-HR models. This stems from a fundamental trade-off between the resolution and variance, which is inherent to the geomagnetic field inverse problem solved here. 

\subsection{Conductivity of the oceanic mantle}

We recall that numerical simulations used for the comparisons with observed ocean tidal magnetic signals were calculated using a global radial conductivity obtained by inversion of the new CI9 $M_2$ tidal signal (see Section \ref{sec:numsim}). Since all constituents sense the same subsurface conductivity, they should tend towards these numerical simulations. To obtain further insights into how satellite-detected ocean tidal magnetic signals for each tidal constituent can sense the lithosphere-asthenosphere system individually, we designed the following parameter study: A 3-D conductivity model of the world ocean and marine sediments as described in Section \ref{sec:numsim} was used. Below these layers, a model with a 80-km thick layer of conductivity $\sigma_{\mathrm{L}}$ followed by a half-space of conductivity $\sigma_{\mathrm{A}}$ was adopted. In reality, the depth to the lithosphere-asthenosphere boundary (LAB) varies with location, but to first order it is controlled by the age of the oceanic crust \cite{naif2021electrical} with an average depth of $\approx 70-80$~km \cite{grayver2016satellite}. 
\begin{figure}[h]
\includegraphics[width=\textwidth]{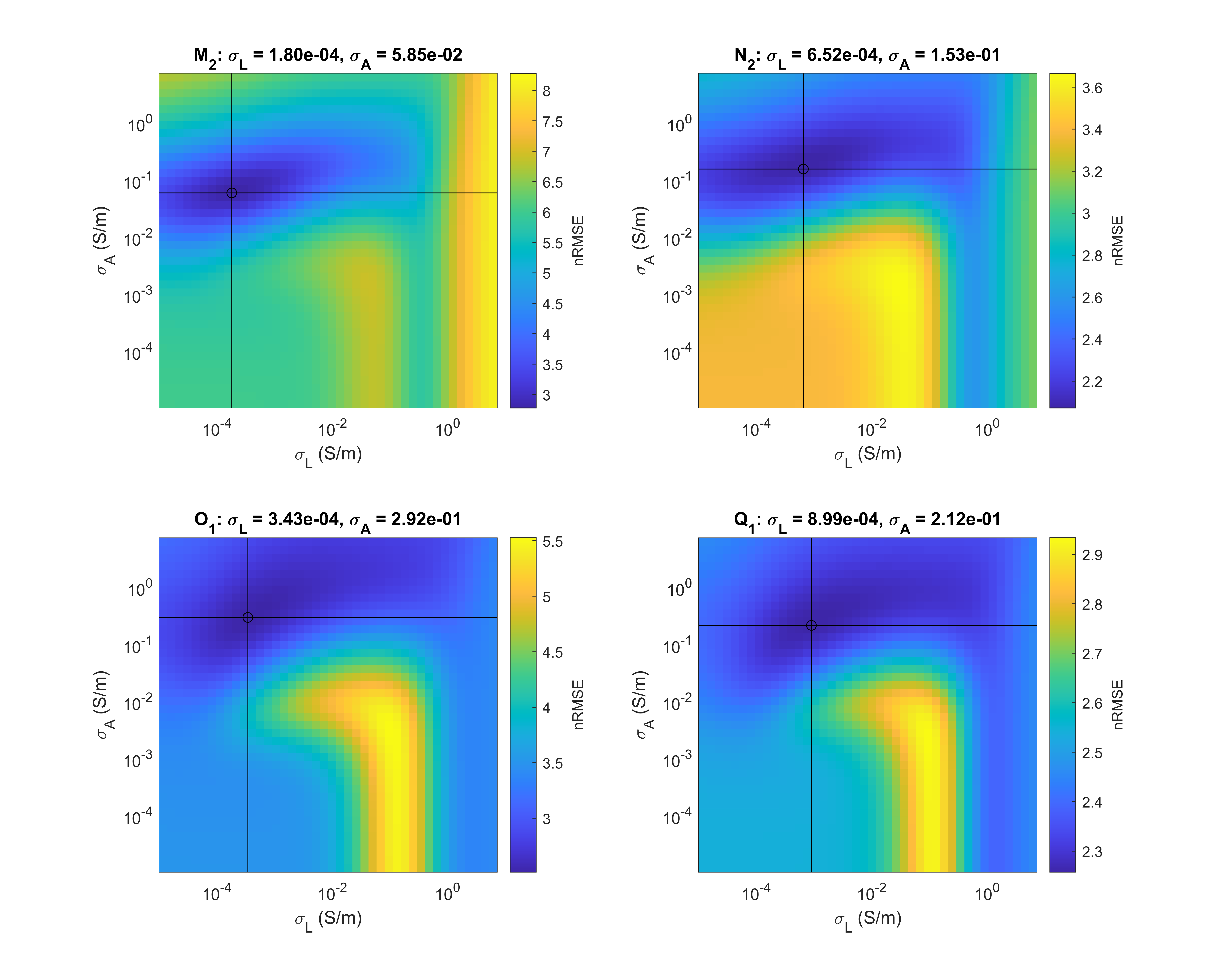}
\caption{Normalized RMS error between simulated and observed signals for various conductivity values of a 80-km-thick lithosphere ($\sigma_L$) and the underlying asthenosphere ($\sigma_A$). Black circle and lines indicate position of the minimum with the corresponding conductivity values specified in the plot titles for each tidal constituent.}
\label{fig:LAB_conductivity}
\end{figure}
The lithosphere and asthenosphere electrical conductivities $\sigma_{\mathrm{L}}$ and $\sigma_{\mathrm{A}}$ were varied in the range $10^{-5}$ to $10^1$ S/m. For every combination of these values, the normalized RMS error between observed and simulated fields at satellite altitude (430~km) was calculated. Standard deviation values for each constituent model from Figure~\ref{fig:Br_stdev}) were used for normalization. The distribution of RMS error values for all tested combinations of the lithosphere-asthenosphere conductivity contrast are presented in Figure~\ref{fig:LAB_conductivity}. Remarkably, all tidal constituents are sensitive to the conductivity gradient across the LAB as is evident from the changes in the normalized RMSE upon variations in the conductivity of the lithosphere against that of the asthenosphere. The steepest slopes in the normalized RMSE around the minimum are seen for the strongest $M_2$ constituent, which has the best signal-to-noise ratio, whereas the opposite is true for the smallest $Q_1$ constituent, although the global minimum is nevertheless clearly visible. Another noteworthy observation is that diurnal tides favour a more electrically conductive asthenosphere, which can be explained since electromagnetic fields at longer periods penetrate deeper and sense the generally more conductive deeper upper mantle \cite{naif2021electrical}. The differences in the retrieved minimum can be explained by different sensitivity patterns, especially between semi-diurnal and diurnal components, as well as by the presence of noise. Additionally, these plots clearly reveal the non-linear nature of the inverse problem and the presence of several local valleys and ridges in the misfit function landscape, motivating the use of advanced non-linear optimization techniques for electromagnetic induction inverse problems.

\section{Discussion and Conclusions}

We presented a new version of ocean tidal magnetic field signals retrieved from satellite magnetic field observations from CHAMP and the ongoing \textit{Swarm} mission. The improved methodology and longer time-series allowed us to substantially improve the quality of the new models reported here, and to increase spatial resolution compared with our previous study. To a large extent these improvements come from using magnetic field gradiometry enabled by the \textit{Swarm} satellite constellation, which suppresses small-scale noise. Systematic comparison with other studies and full 3-D numerical simulations reveals a high level of correlation across all models and shows a trend towards increasing coherency as our time series lengthen and methods advance. 

It is remarkable that all tidal constituents sense an electrically resistive lithosphere and a conductive asthenosphere, corresponding to cold brittle and hotter ductile solid Earth layers, respectively. Although very simplified, this first-order approximation does reflect the overall physical nature of these subsurface layers \cite{naif2021electrical}. Further, since diurnal tides attenuate more slower than semi-diurnal as they diffuse into the upper mantle, they favour a more electrically conductive asthenosphere than semi-diurnal tides. Given the positive conductivity gradient in a normal oceanic asthenosphere \cite{naif2021electrical}, this finding further confirms high quality of the retrieved tidal magnetic signals and their physical nature. 

The detection of the tiny signals from the $Q_1$ tidal harmonic and its sensitivity to a plausible range of electrical conductivities in the upper mantle demonstrates the impressive accuracy of the satellite magnetometers and confirms low calibration errors that can be averaged out very well, at least at the considered frequencies. In this study we were however not able to reliably characterise other tidal constituents, such as $K_1$ or $S_2$. Despite their overall amplitudes being higher than that of $Q_1$ or even $O_1$ (Table~\ref{tab:tides}), their periods (nearly) coincide with the solar daily harmonics. The magnetic field at these periods is dominated by ionospheric fields (see also the discussion in the introduction) even during the quiet conditions considered here \cite{yamazaki2017sq}. It seems justified to extrapolate this argument to ocean tides of lunar and solar origins in general. Specifically, the current trend of improving retrievals of the ocean tidal magnetic signals due to lunar constituents is likely to continue with ongoing high-quality space observations. In contrast, reliable separation of tidal ocean magnetic signals due to solar constituents (or constituents with periods that are very close to the daily period and sub-harmonics) depends largely on how well we can characterize and mitigate the effect of the ionosphere and high latitude currents. The latter challenge is beyond the realm of this study, and will require dedicated efforts in order for progress to be made \cite{finlay2017challenges}.

The model parameterisation given by eq.~(\ref{eq:potential_exp}) targets the stationary harmonic magnetic signal due to a specific tidal constituent of interest. In other word, retrieved signals are independent of variability caused by factors such as seasonal \cite{tyler2017electrical} or long-term changes in the ocean climatology \cite{rhein2013observations}. This approach is preferred for mantle induction applications because the electrical conductivity of seabed and mantle remain essentially constant over the two decades considered here. Since the dependency of tidal magnetic signals on the conductivity of the solid Earth is larger than the component caused by long-term ocean climatology changes, it is important to first characterise the solid Earth part of the signals before trying to retrieve and model putative climatology-driven effects. This appears to be the most consistent way of addressing this non-unique problem. The experience gained with extraction of purely harmonic tidal magnetic signals will certainly be vital when attacking the problem of detecting oceanic magnetic signals due to other forcing mechanisms and detecting climatology-related changes.

The low-altitude phase of the \textit{Swarm} mission during the next solar minimum, a new ESA-Scout mission NanoMagSat consisting of smaller satellites dedicated to measuring the geomagnetic field in orbital configurations complementary to the \textit{Swarm} satellites, and data from the recently launched MSS-1 magnetic survey satellite with its low inclination orbit, will boost the accuracy and resolution of the tidal constituents, unlocking opportunities for addressing new research questions. Consolidation of our methods and modelling tools will ensure that we are able to take full advantage of this next phase of satellite geomagnetism. 

\section*{Data accessibility}
All final models for the four considered tidal constituents, numerical simulation results and scripts to reproduce the figures are archived in a public repository \url{https://github.com/agrayver/magtide}. Swarm data is available at \url{http://swarm-diss.eo.esa.int}. CHAMP mission data is archived at \url{dataservices.gfz-potsdam.de}. Models such as the CHAOS, IGRF, ocean and sediments conductivity are available via cited references. The TPXO-9 model was accessed through the \url{www.tpxo.net/global/tpxo9-atlas} portal. 





\section*{Acknowledgement}
AG is grateful to Gary Egbert for insightful discussions on oceanic tides. We thank Dr.~J.~Baerenzung for providing us with the updated KALMAG models. We used the ChaosMagPy package \cite{clemens_kloss_2024_10598528} for some post-processing and plotting. This work was supported by the Heisenberg Grant from the German Research Foundation, Deutsche Forschungsgemeinschaft (Project No. 465486300) and ESA Swarm DISC project No. 4000109587.



\end{document}